\documentclass{PoS}
\usepackage{lineno}
\usepackage{amsmath}
\usepackage{graphicx}
\usepackage{rotating}
\usepackage{float}
\usepackage{wrapfig}
\usepackage{enumerate}
\usepackage{subcaption}
\usepackage{graphics}
\usepackage{enumitem}

\title{Neutrinos from Primordial Black Hole Evaporation}

\ShortTitle{Neutrinos from PBH evaporation}

\author{

The IceCube Collaboration\footnote{For collaboration list, see PoS(ICRC2019) 1177.}\\
{\itshape \href{http://icecube.wisc.edu/collaboration/authors/icrc19_icecube}{http://icecube.wisc.edu/collaboration/authors/icrc19\_icecube}}\\
E-mail: \email{pdave@gatech.edu, itaboada@gatech.edu}
}

\abstract{

Primordial Black Holes (PBHs) are candidates for dark matter as well as ultra-high energy cosmic rays. PBHs are speculated to exist over a large range of masses, from below $10^{15}$~g to $10^3$~M$_\odot$. Here we search for PBHs with an initial mass of $\sim 10^{15}$~g. Hawking radiation by black holes of this initial mass predicts their evaporation at present time. PBHs are expected to produce copious amounts of high-energy neutrinos and gamma rays right before evaporating. Gamma-ray instruments such as Fermi, VERITAS, HAWC, HESS, and Milagro have conducted searches for evaporating PBHs during their last second to a year of existence. They are able to detect bursts from PBHs in a range of $10^{-3}$ to $0.1$ pc. We present sensitivity to PBH evaporation  using one year of neutrino data by IceCube. In these proceedings, we detail the changes to adapt IceCube's standard neutrino flare search, aka time-dependent point source search, into one that is appropriate for evaporating BHs. These proceedings serve as proof of concept for a first-ever search for evaporating PBHs using neutrinos that can use 10 years of IceCube data.

\vspace{4mm}
{\bfseries Corresponding authors:}
\speaker{Pranav Dave}$^{1}$, Ignacio Taboada$^{1}$\\
{$^{1}$ \itshape Georgia Institute of Technology}

}

\FullConference{36th International Cosmic Ray Conference -ICRC2019-\\
		July 24th - August 1st, 2019\\
		Madison, WI, U.S.A.}

\begin{document}
\section{Introduction}\label{sec:intro}
Hawking proposed \cite{Hawking:1974} that thermal radiation due to quantum gravitational effects near the event horizon of a black hole would cause it to lose mass. The black hole temperature, which determines the instantaneous spectrum of Hawking radiation, and the remaining lifetime of the black hole are a function of the black hole's mass. As detailed later, the temperature of the black hole increases rapidly towards the end of the life of the black hole, leading to a burst in radiation. 

Black holes of mass < $10^{15}$ g would completely evaporate over the age of the universe. Fluctuations in the early universe could form such small black holes. Over the last decade, primordial black holes (PBHs) have been considered to be dark matter candidates. Assuming a narrow initial mass distribution, black holes can only be consistent with a dark matter hypothesis if they are in the ranges $10^{16} - 10^{17}$~g (asteroid mass scale); $10^{20} - 10^{24}$~g (sublunar mass scale) and $10 - 10^3$~M$_\odot$ (intermediate mass black hole or IMBH) \cite{2019arXiv190107803C}. The latter mass range has recently received a lot of interest due to the unexpected detection of tens of solar mass black hole mergers by LIGO/VIRGO \cite{gw150914}. PBHs that have already evaporated  have cosmological consequences, however, they do not contribute directly to currently existing dark matter \cite{2019arXiv190107803C}. For broad initial mass distributions, PBHs with initial mass of $\sim 10^{15}$~g, that are currently evaporating, remain of interest. 

A recent study by Fermi LAT \cite{fermilat:2018} searched for GeV gamma rays from PBHs over their remaining months to a few years of existence, when their mass had been reduced to $6\times10^{11}\,\mathrm{ g}$. Because Fermi LAT is sensitive to PBHs at a distance of $\sim 0.03$pc, the study took into account the proper motion of PBHs. Fermi LAT sets the local density rate for evaporating PBHs at $\dot{\rho} < 7.2\times 10^{3}$~pc$^{-3}$yr$^{-1}$ with 99\% confidence level. This is currently the most constraining density rate limit published on the evaporation of (initial mass) $\sim 10^{15}$~g PBHs. HAWC is expected to reach a similar sensitivity using 5 years of data, but using TeV gamma rays and studying the last $\sim$10~s of the life of the PBH. Milagro, the predecessor of HAWC, placed \cite{Milagro:2015} an upper bound of $\sim 4\times10^{4}\,\mathrm{pc}^{-3}\mathrm{yr}^{-1}$ on the same time scale and energy as HAWC. Imaging Air Cherenkov Telescopes VERITAS \cite{Te_i__2012} and H.E.S.S. \cite{2013arXiv1307.4898G} have also conducted searches.  

IceCube is a cubic-kilometer neutrino detector deployed deep in the ice at the geographic South Pole \cite{Aartsen:2016nxy}. Reconstruction of the direction, energy and flavor of the neutrinos relies on the optical detection of Cherenkov radiation emitted by charged particles produced in the interactions of neutrinos in the surrounding ice or the nearby bedrock. More details about IceCube can be found on proceedings to this conference \cite{dwilliams:2019icrc}. Halzen and Zas \cite{Halzen:1995hu} argued in 1995 that neutrinos from PBH evaporation can place comparable bounds as from bursts of $\gamma$-rays, if the neutrino detector is sensitive to high energies and is kilometer-sized. Furthermore, Halzen and Zas also estimated that such a neutrino detector would be sensitive to PBH evaporation up to a distance of $0.001 - 0.1$~pc. In these proceedings, we use 1 year of track-like events \cite{realtime:2016} used by IceCube's alert system. When a $\nu_\mu$ interacts in ice/rock via the charged current interaction, it produces a muon ($\mu$) which crosses the detector before decaying and leaves a track of light in the detector. Our preliminary results using 1 year of data show that IceCube is most sensitive to a $10^3$~s flares and with a local density rate sensitivity in the range of $3.8\times10^{6}\,\,\mathrm{pc}^{-3}\mathrm{yr}^{-1}$. While this is less sensitive than gamma ray instruments, we should be able use 10 years of data for a significant improvement of sensitivity and this is the first time that a search using neutrinos will be conducted. 

\section{Method}\label{sec:method}
\subsection{Time-averaged emission spectra}
The no-hair theorem postulates that BHs can be described solely by their charge, angular momentum, and mass. Assuming a neutral, non-spinning BH, we use Hawking's \cite{Hawking:1974} instantaneous rate of particle production in the energy range $(E,E+dE)$:

\begin{equation}
    \frac{d^2N}{dEdt} = \frac{1}{2\pi \hbar}\,\frac{\Gamma_s(E,M)}{\exp\left({8\pi ME/ \hbar c^3}\right) - (-1)^{s}}
\end{equation}

Here the mass of the BH is $M$, spin of the emitted particle is $s$, and the absorption co-efficient is $\Gamma_s(E,M)$. We use the averaged absorption cross-section to calculate $\Gamma_s(E,M)$ as in \cite{Halzen:1995hu}. Additionally, a temperature for the BH can be defined as:
\begin{align*}
    kT = \frac{\hbar c^3}{8 \pi GM}
\end{align*}

Due to this inverse relation between the mass and temperature, the BH loses mass and accelerates its temperature rise, and we can use the emission rate to calculate the temperature of the BH as a function of the time remaining to evaporation, $\tau$:

\begin{equation*}
    kT \simeq 7.829\left(\frac{1\,\textrm{s}}{\tau}\right)^{1/3}\,\textrm{TeV}
\end{equation*}

As we show below, IceCube is most sensitive to the remaining $10^3$~s of lifetime, corresponding to a temperature of 783 GeV. We consider four contributions to the neutrino spectra in: (i) direct neutrinos (ii) direct charged pion decay (iii) direct muon decay (iv) hadronization of direct quarks. Channels (ii)-(iv) produce neutrinos once these particles decay. Assuming that the temperature of the BH is much higher than the rest mass of a pion, which is a good approximation over the last 1000~s of lifetime of a BH, the time-integrated emitted particle fluence as a function of remaining lifetime is well approximated by a broken power-law. This is shown in Fig. 1 for the last 1000 and 10~s of lifetime of a PBH. The time integrated fluence between two arbitrary times can be easily calculated by subtracting two broken power law fluences matching the times in questions. Fig. 1 also shows two such fluences. The particle fluence in Fig. 1 has been calculated with a black hole at a reference distance, $d_{ref}$ of 0.01~pc. More explicitly, Fig. 1 shows the particle fluence for a PBH at a distance $d_{ref}$ due to Hawking radiation integrated over time $\tau$ before evaporation:

\begin{equation*}
    \frac{dN_\nu}{dE}(d_\mathrm{ref},\tau,E) = \frac{1}{4\pi d_\mathrm{ref}^2}\int_\tau^0 \frac{d^2N_\nu}{dEdt}\,dt
\end{equation*}

\begin{figure}[ht]
	   	 	\center
	        \includegraphics[width=0.8\textwidth]{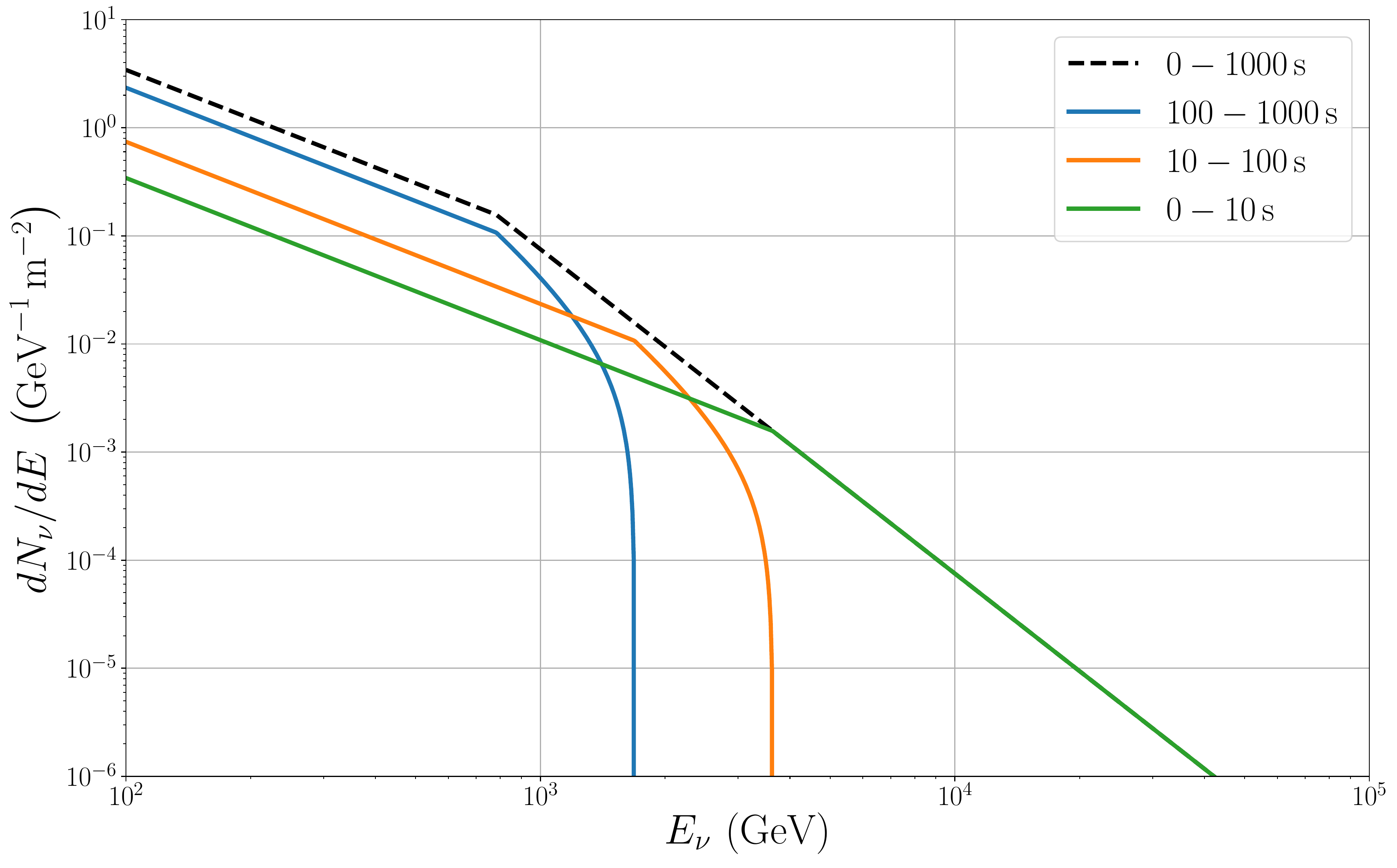}
	        \caption{The approximated time-integrated neutrino fluence as a function of neutrino energy over various intervals from 0 - 1000 s calculated at a reference distance, $d_\mathrm{ref} = 0.01\,\, \mathrm{pc}$. Each solid curve corresponds to one bin of signal events used to calculate sensitivity. The break in the black dashed and green curves correspond to the temperature of the black hole at 1000 s and 10 s respectively.} 
\end{figure}

IceCube's software which is based on \cite{braun2008}\cite{braun2010} only allows the simulation of signals in which the spectrum is time independent (at least within a range of time). Hence we construct neutrino PBH signal using the 3 bins in time shown in Fig. 1.  We are aware that this is not optimal and more bins are needed to capture a typical PBH burst in neutrinos. This will be implemented in the future. 

\subsection{Unbinned likelihood method}

We employ a time-dependent unbinned maximum likelihood method as described in \cite{i3timedep:2015}: 

\begin{equation}
        \mathcal{L} = \sum_i^N \left( \frac{n_s}{N}\cdot S_i + \left(1-\frac{n_s}{N} \right) \cdot B_i   \right),
        \label{eq:eq1}
\end{equation}
where $S_i$ and $B_i$ are the signal and background PDFs for each event. $n_s$ and N are the signal and total neutrino events respectively. The signal PDF includes a temporal, directional, and energy term \cite{i3timedep:2015}. The parameters in the signal PDF are: spectral index of signal events ($\gamma$), mean time of the gaussian signal flare ($T_0$), and the width of the gaussian signal flare ($\tau$). We calculate the test statistic by:

\begin{equation}
    TS = -2 \log \left[\frac{T_0}{\sqrt{2\pi}\hat{\sigma}}\,\frac{\mathcal{L}(n_s=0)}{\mathcal{L}(\hat{n}_s,\hat{\gamma},\hat{\tau},\hat{T}_0)}\right],
     \label{eq:eq2}
\end{equation}
where $\tau$ and $T_0$ are the width and mean of the Gaussian time PDF to be fitted, and $\hat{n}_s$, $\hat{\gamma}$, $\hat{\tau}$, and $\hat{T_0}$ are the best-fit values which maximize the likelihood. $\mathcal{L}(n_s=0)$ is the likelihood of background-only null hypothesis and $\mathcal{L}(\hat{n}_s,\hat{\gamma},\hat{\tau},\hat{T}_0)$ is the maximum likelihood of the signal+background hypothesis. The test-statistic defined this way would follow a $\chi^2$ distribution where the degrees of freedom reflect the number of fit-parameters.

We use 1 year of Gamma-ray Follow Up (GFU) data \cite{realtime:2016} which is dominated by atmospheric muon neutrinos. We can use data to calculate the properties of a background only observation. This is achieved by randomizing, or scrambling, events in time. The scrambled data are obtained by first finding the time intervals when the detector is active. Then random times for events are generated within those allowable intervals. Simulated signal with the method described above can be added, or injected, on top of scrambled data to test the ability of IceCube to detect PBH evaporation. After scrambling the data, we select events within a $\pm 10^\circ$ box in the sky around the injection coordinates. We then maximize the likelihood for those events by adjusting the fit parameters. We find that the Gaussian flare hypothesis is effective in identifying PBH evaporation.  


\section{Results}

As described in \cite{i3timedep:2015}, we calculate IceCube's sensitivity to neutrinos from PBH bursts by calculating the required $n_s$ to attain 90\% of injection TS distribution to be above the median of the background-only TS distribution. Similarly, the discovery potential requires the median of the injection TS distribution to be above the 5$\sigma$ level compared to background. Using a reference distance, $d_\mathrm{ref}$, to the PBH in our fluence calculations, we compute the corresponding number of neutrino events, $n_\mathrm{ref}$, in IceCube by integrating the particle fluence with the effective area $A_\mathrm{eff}(\delta, E)$,  to convert the sensitivity, $n_\mathrm{sens}$, and discovery potential, $n_\mathrm{disc}$,  to a sensitivity, $d_\mathrm{sens}$, and discovery potential, $d_\mathrm{disc}$, distance.

\begin{equation*}
    n_\mathrm{ref} = \int_0^\infty \frac{dN_\nu}{dE}(d_\mathrm{ref},\tau,E) \,A_\mathrm{eff}(\delta,E)\, dE
    ,\quad \quad d_\mathrm{sens} = d_\mathrm{ref}\times \sqrt{\left(\frac{n_\mathrm{ref}}{n_\mathrm{sens}}\right)}
\end{equation*}

\begin{figure}[ht]
	   	 	\center
	        \includegraphics[width=0.8\textwidth]{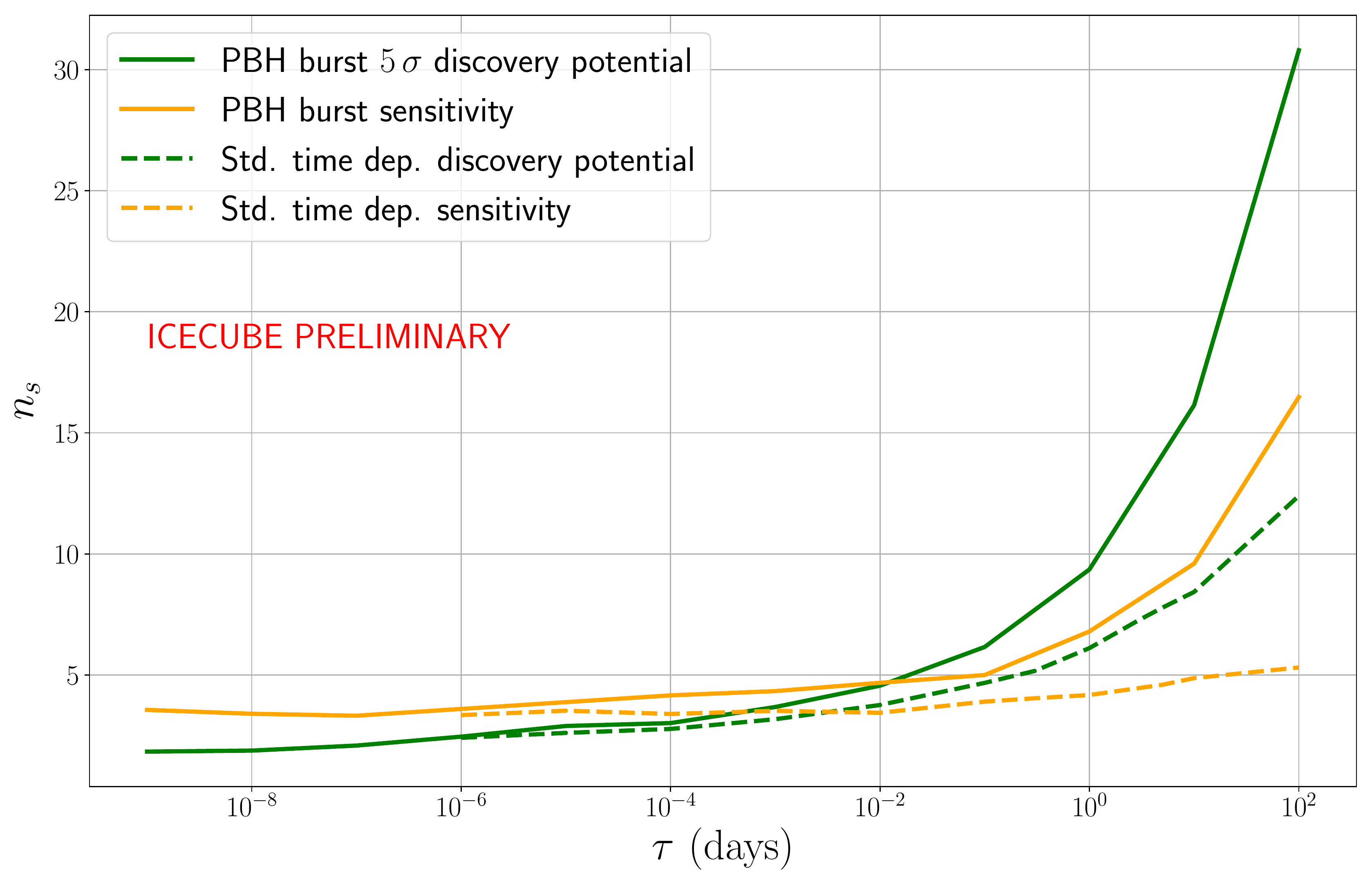}
	        \caption{The number of signal events required for sensitivity and 5$\sigma$ discovery potential for neutrinos from PBH evaporations as compared to the standard gaussian flare search \cite{i3timedep:2015} at declination $\delta = 16^\circ$. $\tau$ is the time remaining to PBH evaporation and the width of the gaussian for the standard time dependent values.} 
\end{figure}

\pagebreak 
Furthermore, we assume a uniform PBH distribution in the local universe and calculate the burst rate density, $\dot{\rho}$, using $n_\mathrm{sens}$ and $n_\mathrm{disc}$ in the following relation:
\begin{align*}
    \dot{\rho} &= \frac{3}{4\pi}\frac{n_\mathrm{sens}}{d_\mathrm{sens}^3}
\end{align*}

\begin{figure*}[t!]
    \centering
    \begin{subfigure}[t]{0.5\textwidth}
        \centering
        \includegraphics[width=\textwidth,clip]{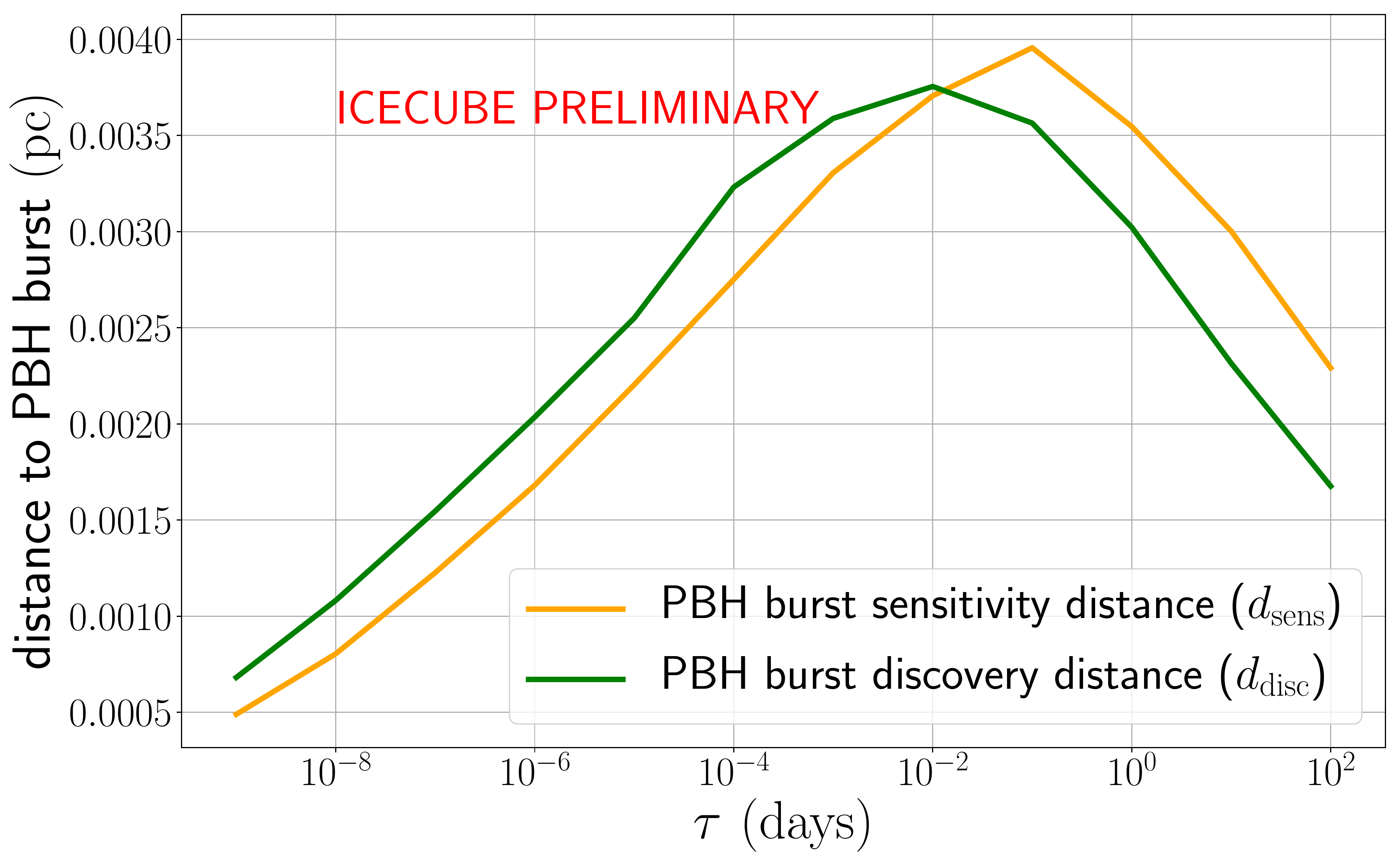}
        \caption{\label{fig:dsens}}
    \end{subfigure}%
    ~ 
    \begin{subfigure}[t]{0.5\textwidth}
        \centering
        \includegraphics[width=\textwidth,clip]{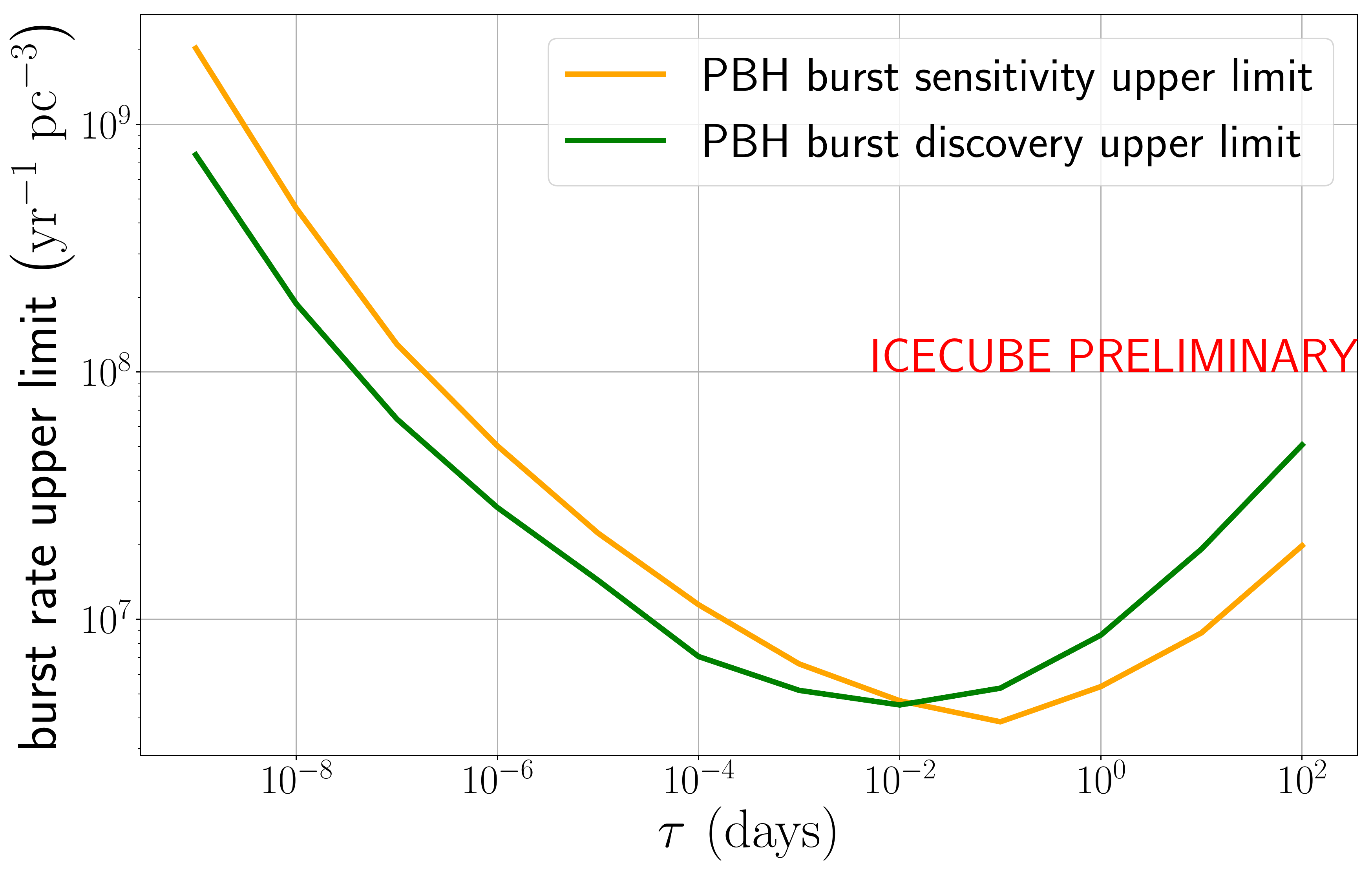}
        \caption{\label{fig:limits}}
    \end{subfigure}
    \caption{\label{fig:results} The left panel (\ref{fig:dsens}) shows the maximum distance to a PBH burst required to achieve 90\% median upper limit sensitivity and 5$\sigma$ discovery potential. The right panel (\ref{fig:limits}) converts the distances in the left panel to burst rate density.}
\end{figure*}
\section{Discussion}
We described a method to search for evaporating PBHs using neutrinos and IceCube data. 
Our preliminary results using 1 year of IceCube data indicates that we would be most sensitive to the final $10^3$ seconds. 
Evaporating primordial black holes would have been reduced from an initial mass of $\sim 10^{16}$~g to  $\lesssim 10^{10}\,\mathrm{g}$ by the time they are detectable by IceCube. The temperature of the PBH to which IceCube is sensitive is $\gtrsim 0.77$ TeV which makes sense since IceCube's sensitivity drops significantly for lower energies and the background drops drastically at higher energies. The sensitivity results in $3.8\times10^{6}\,\mathrm{pc}^{-3}\,\mathrm{yr}^{-1}$ at this flare duration. As a direct comparison, Fermi LAT \cite{fermilat:2018} using GeV gamma rays from $\sim 6\times10^{11}$~g PBHs place their best limits at $7.2\times10^{3}\,\mathrm{pc}^{-3}\,\mathrm{yr}^{-1}$. This is the first study done with neutrinos using the highest energy range ever. We will be able to improve the sensitivity of the search using existing 10 years of IceCube data. Because IceCube software can only simulate time independent spectra, we provided a new method to describe the intrinsically time-dependent PBH spectrum. This is currently limited to 3 bins in time over the remaining life of the PBH, but we should be able to improve on this. 
\bibliographystyle{ICRC}
\bibliography{references}
\end{document}